# Scaling properties of neuronal avalanches are consistent with critical dynamics

*Dietmar Plenz[1] & Dante R. Chialvo[2]*

[1] *Section on Critical Brain Dynamics,
National Institute of Mental Health, Porter Neuroscience Research Center,
Rm 3A-100, 35 Convent Drive, Bethesda, MD 20892, USA
plenzd@mail.nih.gov*
[2] *Department of Physiology, Feinberg Medical School,
Northwestern University, 303 East Chicago Ave. Chicago, IL, 60611, USA.
d-chialvo@northwestern.edu*

**Complex systems, when poised near a critical point of a phase transition between order and disorder, exhibit a dynamics comprising a scale-free mixture of order and disorder which is universal, i.e. system-independent** (1-5)**. It allows systems at criticality to adapt swiftly to environmental changes (i.e., high susceptibility) as well as to flexibly process and store information. These unique properties prompted the conjecture that the brain might operate at criticality** (1)**, a view supported by the recent description of neuronal avalanches in cortex *in vitro* (6-8), in anesthetized rats (9) and awake primates (10), and in neuronal models (11-16). Despite the attractiveness of this idea, its validity is hampered by the fact that its theoretical underpinning relies solely on the replication of sizes and durations of avalanches, which reflect only a portion of the rich dynamics found at criticality. Here we show experimentally five fundamental properties of avalanches consistent with criticality: (1) a separation of time scales, in which the power law probability density of avalanches sizes $s$, $P(s) \sim s^{\alpha}$ and the lifetime distribution of avalanches are invariant to slow, external driving; (2) stationary *P(s)* over time; (3) the avalanche probabilities preceding and following main avalanches obey Omori's law** (17, 18) **for earthquakes; (4) the average size of avalanches following a main avalanche decays as a power law; (5) the spatial spread of avalanches has a fractal dimension and obeys finite-size scaling. Thus, neuronal avalanches are a robust manifestation of criticality in the brain.**

The spontaneous interaction of hundreds of thousands of neurons in the cerebral cortex gives rise to a bewildering variety of spatio-temporal activity patterns. A fundamental question in neuroscience is to understand the functional meaning of such pattern variety. In that direction, recent work (6-10, 19, 20) has shown that in superficial layers of cortex, spontaneous neuronal activity patterns emerge in the form of "neuronal avalanches". These are highly diverse bursts of activity exhibiting scale-free features both in space and time. The sizes of avalanches are distributed according to a power-law with an exponent of -3/2 and their durations according to a similar law with an exponent of –2. Avalanches represent a critical balance of excitation and inhibition in the cortex; it is typical that during the course of an avalanche the number of participating neurons neither grows nor substantially decays, such that the branching ratio of the activity stays close to unity, as in other critical systems (21). Neuronal network simulations (11-16) already have replicated quantitatively the observed scale-free distribution of avalanche sizes, including the formulation of a realistic self-organized critical scenario (22). Overall, these results suggest that neuronal avalanches indicate a critical dynamics of the cortex (6, 8). Such a dynamics would allow cortical networks to gain universal principles found at criticality that are beneficial in numerous aspects such as input processing (19, 23), information transfer (8), and the ability to generate diverse, but ordered internal states enhancing information storage and system adaptability (1).

Despite the importance of these advances, the breadth of this conjecture is limited because the underlying avalanche size and lifetime distributions are insufficient to determine the dynamical origin of the avalanche process, since similar distributions can arise from a variety of model classes including some without critical dynamics. Importantly, the distributions do not describe the temporal evolution of successive avalanches in space and time nor do they provide insight into the ordering of avalanches of different sizes. Here, we present five new key signatures of the avalanche dynamics all of which are consistent with critical dynamics of the cortex.

**Results**

Neuronal electrical activity is recorded from mature organotypic cultures of rat somatosensory cortex (20) where neuronal avalanches spontaneously emerge in superficial layers. Briefly, the data consist of voltage time series detected by a square array of 60 equally spaced (0.2 mm) electrodes (Fig. 1A). As previously described (8), a fast, transient voltage fluctuation in the local field potential (LFP) exceeding –3SD of the noise indicates neuronal activations in the vicinity of each electrode. The peak time and peak amplitude of each suprathreshold LFP event (nLFP) is measured and saved for further analysis (Fig. 1B). As shown in Figure 1C, avalanches are identified as successive nLFPs on the array at temporal resolution $\Delta t$ allowing for the definition of size (both in number of electrodes and voltage) and lifetime for individual avalanches, as well as waiting time and quiet time between successive avalanches.

**Separation of time scales between the triggering and the spread of neuronal avalanche activity**

Our experiments (5 – 10 hrs each) comprise two well-controlled conditions designed to investigate the dynamical properties of neuronal avalanches. In the first one, the culture is submerged under steady laminar flow in ACSF (8). No external manipulation is being made to the culture and data collected should be interpreted as spontaneous, non-driven activity (nd; n=7 cultures). In the second, driven condition (d, n=6 cultures), cultures are submerged in culture medium inside an incubator (20). The entire setup is periodically tilted (with a cycle $\tau = 200$ sec.) such that the fluid bathing the tissue is slowly removed and metabolic conditions for the culture are gradually changed resulting in rhythmic changes of the spontaneous, average rate of nLFPs per electrode by ~500% (0.2±0.2 Hz to 1.0±0.5 Hz). The two top panels in figure 1D are examples of typical nLFP raster plots for the two conditions.

First, we address how temporal driving affects the statistical properties of the avalanche dynamics. Visual inspection of activity on the arrays reveals a similar temporal organization of nLFPs for the driven and non-driven condition, where large nLFP bursts are often followed by successively smaller bursts for temporal resolutions <100 s (Fig. 1D, *bottom*), suggesting similar avalanche dynamics. Indeed, for the non-driven condition, avalanche sizes scale as $P(s) \sim s^{\alpha}$ with $\alpha \approx -3/2$ (Fig. 1E, *left*; $\alpha^{nd} = -1.52 \pm 0.2$) and avalanche lifetimes obey a truncated power law as reported previously (8) (Fig. 1F). These statistics remain unchanged (Fig. 1E, *right*; $\alpha^{d} = -1.54 \pm 0.2$; $p > 0.05$; Fig. 1F) despite the slow periodic modulation introduced by the driving as seen in the autocorrelation of the rate of avalanches (cf. Fig. 1E, *insets*). Similarly, for relatively short time scales, consecutive inter-avalanche intervals, i.e. waiting times *t*, are always independent of preceding waiting times regardless of the presence or absence of driving. This is demonstrated by the fact that the expected value of $t_{i+1}/t_{avg}$, is constant and independent of the value $t_i/t_{avg}$ where $t_{avg}$ is the average waiting time (Fig. 1G). The driving, however, results in relatively long waiting times ($t > t_{avg}$) to be followed, on average, by long waiting times. Importantly, this dependency differs significantly from a simple, decaying function naively expected for oscillatory rates (discussed in (24)), and instead is almost identical to the statistics of waiting times between consecutives earthquakes calculated from the Southern California Seismic Network (SCSN) catalog (25).

The results above indicate that there is a separation of time scales between the process involved in the avalanche initiation and the dynamics of the avalanche itself. This separation of time scales is a hallmark of several models of criticality where the avalanche statistics is unaffected by a (slow enough) external forcing (1-3) and supports recent reports of avalanches embedded in nested brain oscillations at θ-, β- and γ-frequencies (9).

**Stationarity in the generation of neuronal avalanches**

Next we demonstrate that, despite the large fluctuations in the rate and size of avalanches, the power law distribution of avalanche sizes *s* is stationary. This is an important prerequisite to further investigate any critical mechanism, given the fact that trivial non-stationary processes could easily produce such power laws. Figure 2 introduces results for one representative, non-driven experiment where rate fluctuations in avalanche initiation for avalanches of size $s \geq s_0$ are analyzed. We show that, within the range where the power law scaling is valid, the rate of avalanches of a given size is scale invariant. This is demonstrated by the fact that, after appropriately rescaling of the rate $\lambda$ by the mean rate $\lambda_0$ for avalanches $s \geq s_0$, all time series and rate distributions collapse (Fig. 2A - C). The collapse is obtained for non-driven networks (Fig. 2D) as well as for driven networks, when the window for the rate calculation is made longer than the driving period (Fig. 2E). Figure 2F summarizes this scaling relationship for all experiments, denoted by the proportionality between the mean and variance of $\lambda_0$. The stationarity demonstrated here for the initiation of avalanches of all sizes is in line with predictions from criticality and, at the same time, excludes alternative models for avalanche generation such as trivial mixing of random processes with different size dependent rates.

**Temporal correlations between neuronal avalanches are governed in size and rate by power laws: the Omori law**

Now we turn to investigate correlations between avalanches separated by relatively short time intervals. This has been studied at length in a variety of critical systems (1, 26) because of the distinct character of the temporal correlations at the critical state as well as its importance for predicting future events such as earthquakes, solar flares, and forest fires (1). Specifically, we identify all avalanches (N = 1,000 – 25,000) over the entire experiment within a specific range of sizes $s_i \leq s < s_{i+1}$. Then, we compute the probability density of all avalanches before and after such events of defined size as a function of time. The results in figure 3A show for a single network that the probability of avalanches before and after any given avalanche at time $t_0$ follows a power law $P(t) \sim |t - t_0|^{-1}$ with unity slope for time intervals up to about a second, that is about 1 – 2 orders of magnitude longer than the lifetime of most avalanches (see Fig. 1F). This relation holds irrespective of the size of the trigger avalanche and is also not affected by external driving, again, demonstrating a separation of time scales consistent with critical dynamics (Fig. 3B). The scaling is very little affected by shuffling avalanche sizes or quiet times (Suppl. Info Fig. S2A), but is readily destroyed, if quiet times are replaced with intervals from a uniform distribution (Fig. 3B). The power law statistics surrounding trigger avalanches of particular sizes is analogous to the Omori law (17, 18), which captures the dynamics of fore- and aftershocks near a main earthquake (27-29) and in critical avalanche models (30).

The average size of an avalanche is also a function of the time elapsed after a trigger avalanche. It decays according to a power law with slope −1 (Fig. 3C) and the dependency is destroyed when avalanche sizes are shuffled (Suppl. Info Fig. S2B). The relationship is independent of the minimal size $s_0$ of the trigger avalanche, though in general, large trigger avalanches are followed by smaller 'aftershocks', whose sizes decay as a power law during hundreds of ms.

**The fractal dimension in the spread of neuronal avalanches**

To be consistent with a critical scenario, the temporal scale invariance of the avalanches on the array should be also accompanied by spatial scale invariance. Importantly, in a critical system, all sites involved in an avalanche form a fractal object with a given fractal dimension $d_F$. We demonstrate this property for neuronal avalanches using a finite-size scaling analysis. Specifically we determine how the mean rate $\lambda_0$ for avalanches $s \geq s_0$ scales with minimal avalanche size $s_0$ and spatial dimension $L$ (see Fig. 4A). We first show how $\lambda_0$ changes with $L$ for a given $s_0$ (Fig. 4B; *inset*). A collapse of these functions obtained for different size $s_0$ is achieved for the scaling $L \rightarrow L \cdot s_0^b$ with $b^{nd} = -0.56 \pm 0.04$ and $b^d = -0.6 \pm 0.03$ ($p > 0.05$). The scaling region for which $\lambda_0 \sim (L \cdot s_0^b)^\gamma$ can be observed for $L \cdot s_0^b > 0.03$ and $\gamma \cong 1.1$ for the single, driven network in figure 4B. Conversely, for a fixed spatial size $L$, we calculate the decrease in $\lambda_0$ as $s_0$ increases (Fig. 4C, single non-driven network; *inset*). Collapse of these functions is obtained with the scaling $s_0 \rightarrow s_0/L^{d_F}$. Note that in general $d_F = 1/b$, but now $d_F$ is an estimate of the spatial dimension of the avalanche process. For a compact, two-dimensional process, such as the spreading of a wave on a surface, $d_F = 2$. Instead, for the spread of avalanches on the array, we obtain the fractal dimension $1 < d_F < 2$, with $d_F^{nd} = 1.6 \pm 0.1$ and $d_F^d = 1.76 \pm 0.1$ ($p>0.05$). An equivalent analysis, when calculating avalanche size in number of electrodes and expressing $L$ in multiples of the inter-electrode distance of 200 µm, yields similar values ($d_F^{nd} =1.71 \pm 0.2$; $d_F^d =1.71 \pm 0.1$) and allows for an overplot of all experiments (Fig. 4D). For comparison notice that Albano (31) found $d_F = 1.8$ for the activity spread in forest fire models at criticality in two dimensions, which is in good agreement with our estimation, if one consider that we arrived at our figures trough a scaling argument. Probably, the agreement can be even better if differences in network topology (32) and finite size effects are considered.

**Discussion**

The analysis presented here uncovered five quantitative novel aspects of the dynamics of neuronal avalanches all consistent with critical dynamics, including 1) the separation of times scales between the triggering and the avalanching event itself, 2) the demonstration of stationarity in its size distribution, 3) power laws describing the temporal clustering of avalanches before and after relatively large ones, resembling the Omori law, 4) a power law decay for the size of avalanches following a relatively big one, 5) a fractal dimension, which scales the spatial spread of avalanches for any given avalanche size and area. While future technological advances ultimately might allow for the identification of critical exponents (5) as well as the universality class of this dynamics (30), thus unambiguously proving the existence of criticality, this is by far an unrealistic objective. Our results provide solid experimental evidence that the scale-invariant properties of neuronal avalanches arise in a cortical network at criticality. They also exclude alternative arguments such as the combination of various statistical processes and overall provide precise guidance to advance current models (11-16, 22) for this particular cortical dynamics, which so far successfully replicate only the size and lifetime distributions of neuronal avalanches.

The implications of the present results can be extended to the dynamics of the brain at larger scales, as shown by a recent study on human brain network synchronization (33) using MRI and magnetoencephalographic techniques. By measuring the statistics of the periods of phase-locking between multiples brain sites, the authors were able to demonstrate the existence of power law scaling for the synchronization metrics. The occurrence of these prolonged periods of phase-locking interrupted by rapid changes in the state of global synchronization, were interpreted as "analogous to the neuronal avalanches previously described in cellular systems" (33) indicative of criticality.

The findings reported here change our conception of how cortical networks organize their intrinsic activity. An avalanche is not an isolated event, but rather its specific value in time, space, and size is part of an elementary organization of the dynamics that extends over many orders of magnitudes for all three dimensions. This 'superstructure' which guides prediction of future events as well as reconstruction of past activity is described by three basic power laws that scale avalanche size ($\alpha$), avalanche time (Omori-law), and avalanche spread ($d_F$). We propose that the scaling laws uncovered here for neuronal avalanches open a novel framework to understand cortex function in terms of critical neuronal activity, much in the same way as decades ago, when the discovery of distinct oscillations shaped our understanding of cortex function in terms of phase and synchrony.

**Material and Methods**

**Experimental setup.** Coronal slices from rat dorsolateral cortex (postnatal day 0–2; 350 μm thick) were attached to a poly-D-lysine coated 60-microelectrode array (MEA; Multichannelsystems, Germany) and grown at 35.5 °C in normal atmosphere in standard culture medium without antibiotics for 4 – 6 weeks before recording. Two sets of cultures were used for the study. The first set consisted of avalanche activity from cortex-striatum-substantia nigra triple cultures or single cortex cultures as reported previously (8). In short, spontaneous avalanche activity was recorded outside the incubator in standard artificial cerebrospinal fluid (ACSF; laminar flow of ~1 ml/min) under stationary, non-rocking conditions for up to 10 hrs (nd; non-driven). The second set of cultures was recorded inside an incubator in 600 µl of culture medium under sterile conditions for at least 5 hrs (20). The head stage with the MEA was rocked between ±75º with a sinusoidal trajectory and a pause of 10 s at the steepest angles before reversing direction (cycle time $\tau \cong 200$ s). The exposure to atmosphere at the steepest trajectory angles triggered transient neuronal activity increases (d; driven). The spontaneous local field potential (LFP) was sampled continuously at 1 kHz at each electrode and low-pass filtered at 50 Hz. Negative deflections in the LFP (nLFP) were detected by crossing a noise threshold of −3 SD followed by negative peak detection within 20 ms. nLFP times and nLFP amplitudes were extracted.

**Neuronal avalanches.** Neuronal avalanches were defined as spatiotemporal clusters of nLFPs on the MEA (for details see (20)). In short, a neuronal avalanche consisted of a consecutive series of time bins with width $\Delta t$ that contained at least one nLFP on any of the electrodes. Each avalanche was preceded and ended by at least one time bin with no activity (Fig. 1C). Without loss of generality, the present analysis was done with width $\Delta t$ individually estimated for each culture from the average inter nLFP interval on the array (20) at which the power law in avalanche sizes $s$, $P(s) \sim s^{\alpha}$ yields $\alpha \cong -3/2$. $\Delta t$ ranged between 3 – 6 ms for the two sets of cultures. Avalanche size was defined as the sum of absolute nLFP amplitudes (µV) on active electrodes or simply the number of active electrodes (n). Size distributions were obtained using logarithmic binning for sizes expressed in µV.

**Scaling of avalanche rate for spatial extent and minimal avalanche size.** For the 8x8 MEA (corner electrodes missing) with an interelectrode distance of $\Delta L = 0.2$ mm, the spatial scale $L$ ranged from 0.2 to 1.6 mm, which corresponded to squared 1x1 to 8x8 $(\Delta L)^2$ sub-regions on the array respectively (Fig. 4A). Thus, for a given dataset and e.g. $L = 0.8$ mm, avalanches were calculated from a total of 16 different, partially overlapping 4x4 subregions and the results were averaged. The lack of corner electrodes on the MEA restricted the maximal number of electrodes to 60 electrodes. It also reduced the number of electrodes for corner subregions, however, with negligible effect on the results. Avalanche rate $\lambda_0$ as a function of spatial dimension $L$ and minimal size $s_0$ was calculated by dividing the number of avalanches with sizes above threshold $s_0$ that occurred within an $LxL$ subregion by the total length of recording time (usually between 5 – 10 hr). $\lambda_0$ was determined for values of $L$ from 0.2 to 1.6 mm in linear steps of $\Delta L = 0.2$ mm, and $s_0$ values from 3 – 4000 µV in logarithmic steps or 1 to 60, the number of electrodes on the array. For scaling in $L$, $L \rightarrow L \cdot s_0^b$ and $s_0$, $s_0 \rightarrow s_0/L^{d_F}$, the corresponding exponents $d_F$ and b were determined by minimizing the absolute error between the collapsed, mean function and individual functions for b = [–0, 0.05, …, –1] and $d_F$ = [0.5, 0.6, …, 3]. Minimum error was identified based on the fit of a quadratic function. After determining optimal values for $d_F$ and b, scaled, average rate function were calculated using logarithmic binning.

Autocorrelation functions (ACFs) were calculated from the time series of nLFPs for each electrode and averaged. Statistical significance was obtained at $p < 0.05$.

**ACKNOWLEDGEMENTS**

We thank Drs. W. Shew and S. Pajevic for their detailed comments on an earlier version of the manuscript. This work was supported by DIRP National Institute of Mental Health. DRC is supported also by NINDS.

# FIGURES

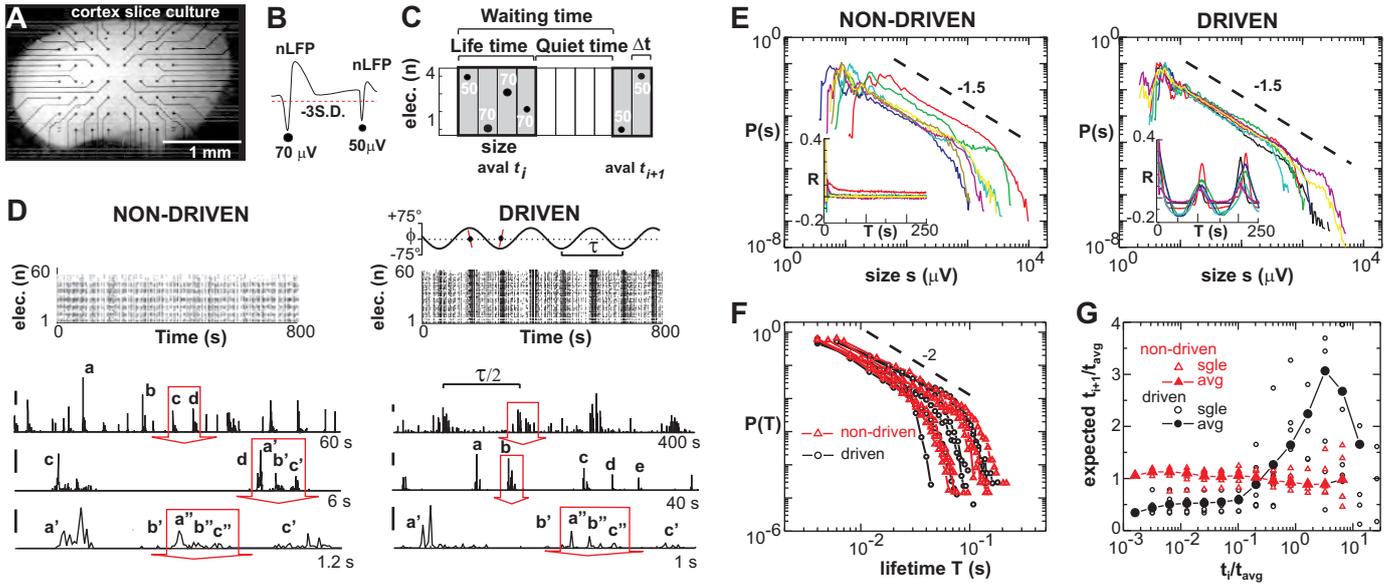

**Fig. 1. Hierarchical organization of neuronal avalanches in space and time demonstrates separation of time scales characteristic of criticality.**
(*A*) Light microscopic image of a coronal, 6-layer cortex slice grown on a 60-microelectrode array (MEA; black dots and leads). (*B*) Significant negative deflections in the local field potential (nLFP; <–3SD of the noise) are indicative of local synchronized neuronal activity in the network (nLFP peak time and absolute peak amplitude indicated by filled dot size and number in µV). (*C*) Neuronal avalanches (*gray*) are defined on a sequence of consecutively active time bins of width *Δt*, bracketed by at least one time bin with no activity (sketch on a 2x2 array). Avalanche's size, lifetime, absolute nLFP peak amplitudes in µV, and the waiting as well as quiet time between avalanches are defined in the cartoon. (*D*) *Top*: Temporal raster display of nLFPs (*dots*) for each electrode (row; single experiment). *Left*: non-driven condition under steady laminar flow. *Right*: Driven condition, in which slow rocking (T ≅ 200 sec cycle time; *Top*) inside an incubator induces rhythmic changes of the nLFP rate at extreme angles (*red lines*) with period τ/2. *Bottom*: Non-driven and driven networks show a similar hierarchical organization of nLFPs clusters in time for time scales <200s. Integrated nLFP activity on the array (*Δt* = 6ms; n = 60 electrodes) forms temporal clusters where many large events are successively followed by smaller events (a –> b –> c –> d –> …). Large events are composed of self-similar hierarchical events at higher temporal resolution (e.g. *left*: d –> a', b', c'; b' –> a", b", c"). Three different time scales (*top*, *middle*, b*ottom*) plotted for driven and non-driven conditions. *Red arrows* indicate subsequently enlarged time periods. (*E*) Neuronal avalanche size *s* scales according to a power law P(*s*) ~ $s^\alpha$ with slope α = – 1.5 (*broken line*). The power law is preserved in the driven condition demonstrating separation of fast time scale of avalanche generation from slow time scale of external driving (Individual size distributions for n = 7 non-driven cultures (*left*) and n = 6 driven cultures (*right*; >20,000 avalanches each). *Insets:* Corresponding autocorrelations for nLFPs averaged over all electrodes demonstrating absence of significant correlations for times >10 s in the non-driven condition (*left*) and strong temporal correlations induced due to the slow driving (*right*). (*F*) Avalanche lifetimes are similar for non-driven (*red*) and driven networks and rarely exceed 30 – 100 ms (*broken line*: slope of –2). (*G*) The expected waiting time $t_{i+1}$ to the next avalanche does not depend on the preceding waiting time $t_i$. External driving only affects the relatively long intervals (data normalized by the average waiting time $t_{avg}$). Filled symbols correspond to the average for the non-driven (*triangles*) and driven (*circles*) condition calculated from individual experiments (*open symbols*).

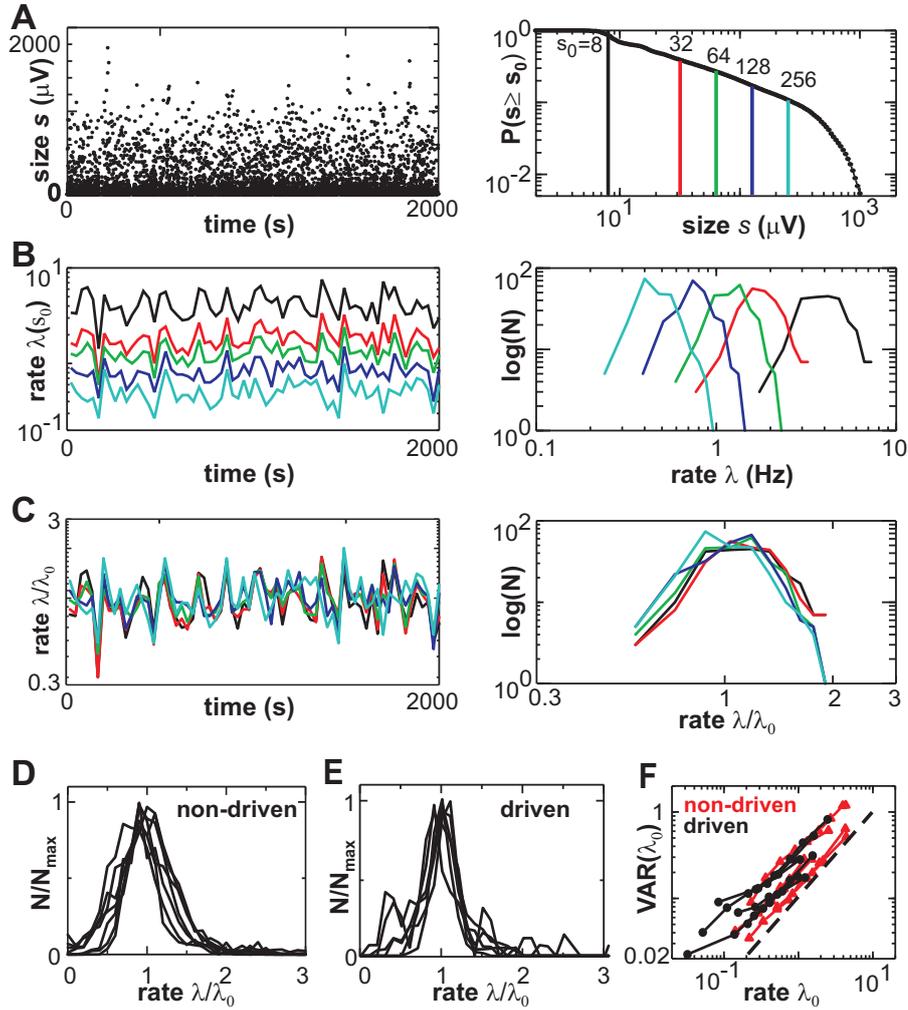

**Fig. 2. Despite large fluctuations in avalanche rate, the size distribution is stationary.**
(*A*) Example time series (*left*) of consecutive avalanche sizes $s \geq 8$ µV for a single non-driven network and its cumulative density distribution (*right*). Size thresholds $s_0$ (*numbers*) used in *B* and *C* indicated in color. (*B*) Time series of avalanche rate $\lambda(s_0)$ for sizes $s \geq s_0$ (*left;* 30 s window) and their corresponding rate histograms (*right*) for 5 different $s_0$. $N$: number of rates observed. (*C*) All $\lambda(s_0)$ time series, when normalized by the corresponding mean rate $\lambda_0$, collapse into a unique series (*left*) and rate histogram (*right*) demonstrating the stationary of the power law distribution of avalanches sizes. (*D*) Similar collapse is demonstrated in the rate distributions for all non-driven and (*E*) driven (window size: 200 s) networks. Average histogram after collapse normalized by the maximal number of rate observations, $N_{max}$, plotted for each network. (*F*) The proportionality between the variance (VAR) and corresponding mean rate $\lambda_0$ for all networks, summarize the generality of the results shown in A-E (*broken line*: slope of 1).

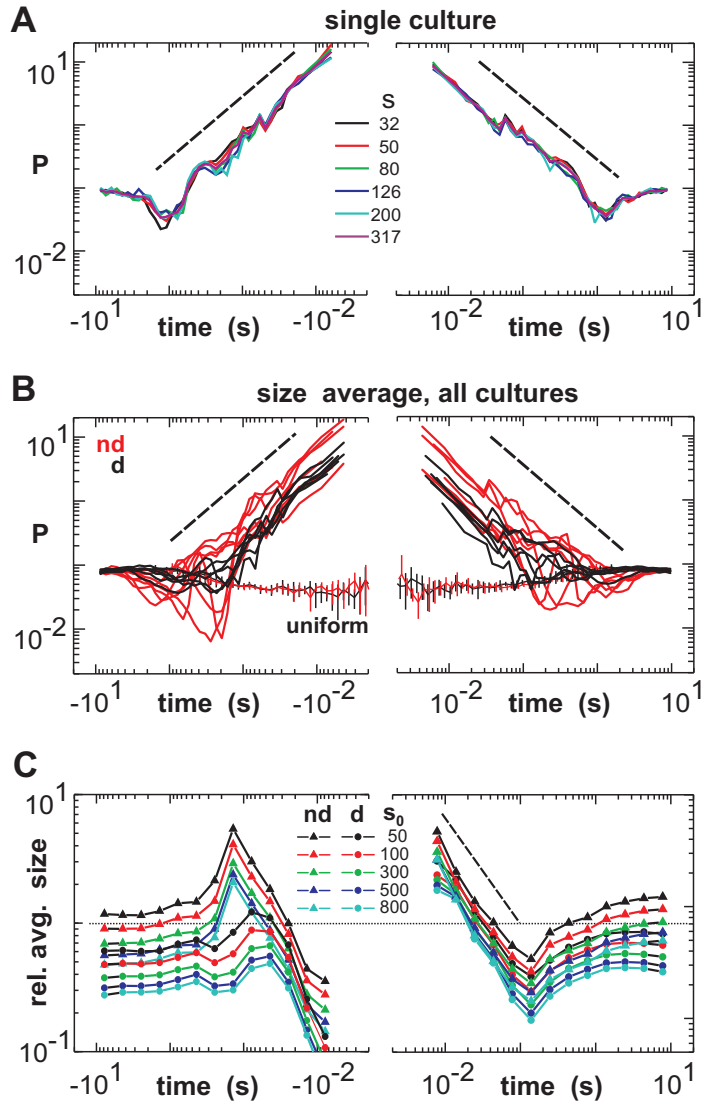

**Fig. 3. The dynamics of events preceding and following a neuronal avalanche obey scaling laws previously described for earthquakes and critical systems.**
(*A*) The frequency of avalanches occurring before (*foreshock*) or after a main avalanche (*aftershock*) follows power laws with unity slope similar to Omori's law for earthquake aftershocks. The equivalent to the Omori law is independent of main avalanche size, as shown by the super imposed plots of the shock probability leading to (*Fore, top*) or following (*After, bottom*) main avalanches of different size *s* (results for a *single non-driven network*). (*B*) Probability for each non-driven (nd, *red*) and driven (d, *black*) network as a function of time averaged over up to 7 main avalanche sizes *s* within the respective power law regime of each network (s = [32, 50, 80, 126, 200, 317, 502, >600 µV]). We note that in order to reduce the effect of lifetimes close to the mainshock, aftershock functions are computed starting at the end of the mainshock, while foreshock functions are measured from the end of foreshocks. Similar results are obtained when considering waiting times only (Suppl. Info Fig. S1). *Uniform quiet*: null hypothesis computed by replacing quiet times with those drawn from a uniform distribution for each network and averaging over driven and non-driven condition respectively. (*C*) Avalanche size decays in a sequence according to a power law both for non-driven (*triangle*) and driven (*circle*) networks in the first 100 - 200 msec. Mean size normalized by the mainshock size plotted as a function of time before and after an avalanche of five different minimal size $s_0$. Note that the decay is independent of main avalanche size (see also Suppl. Info Fig. S2B).

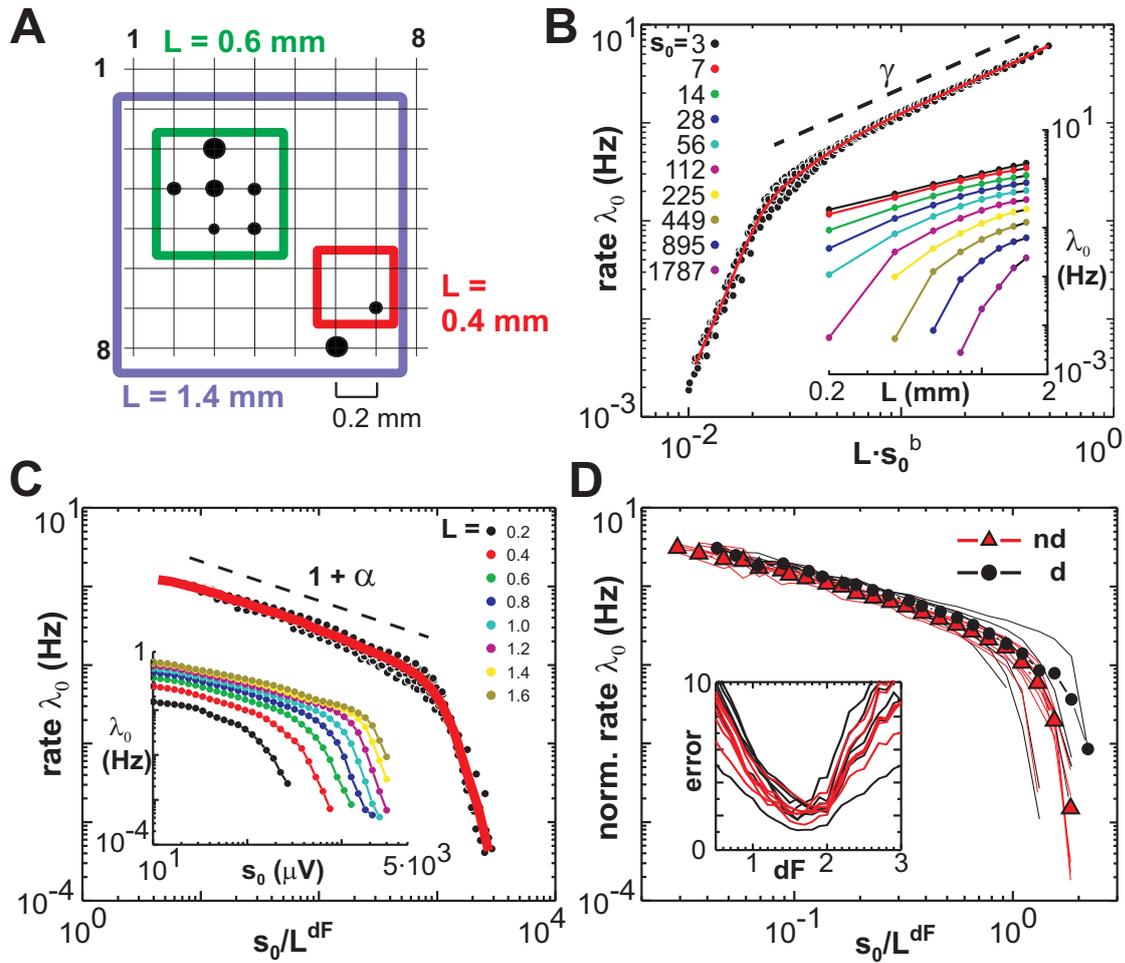

**Fig. 4. The fractal spread of neuronal avalanches**.
(A) Sketch of the spatial occupancy from the beginning to the end of a single avalanche (nLFP: *dots*) on the full grid ($L = 1.8$ mm) and in the observation of subregions $LxL$ with $L = 0.4$, 0.6, and 1.4 mm respectively. (B) Collapsed distribution and average (*red line*) describing the increase of $\lambda_0$ with increase in spatial extent $LxL$ for any given $s_0$ using the scaling $L \rightarrow L \cdot s_0^b$ and $b = -0.54$ (minimal scaling error; *single driven network*). *Inset*: Corresponding non-scaled functions $\lambda_0(L)$ at given $s_0$. (C) Collapsed function for the decrease in $\lambda_0$ with increase in minimal avalanche size $s_0$ and any spatial extent $LxL$ using the scaling $s_0 \rightarrow s_0/L^{dF}$ and $d_F = 1.8$ (minimal scaling error; *single non-driven network*). *Inset*: Corresponding non-scaled functions $\lambda_0(s_0)$ at given $L$. (D) Summary in data collapse for non-driven (*black*), driven (*red*) networks and corresponding averages using the scaling approach in C for avalanche sizes $s$ and threshold $s_0$ calculated in number of electrodes, i.e. active sites. $L$ is expressed in multiples of interelectrode distance $\Delta L$. *Inset*: Corresponding error for each network to minimize collapse as a function of $1 < d_F < 2$ (mean $1.7 \pm 0.1$).

# Reference List

1. Bak, P. (1996) *How Nature Works: The Science of Self-Organized Criticality* (Copernicus Books, New York).

2. Jensen, H. J. (1998) *Self-organized criticality* (Cambridge University Press.

3. Sornette, D. (2000) *Critical Phenomena in Natural Sciences* (Springer Verlag, Berlin).

4. Frette, V., Christensen, K., Malte-Soerenssen, A., Feder, J., Joessang, T. & Meakin, P. (1996) Avalanche dynamics in a pile of rice *Nature* **379**, 49-52.

5. Stanley, H. E. (1971) *Introduction to phase transitions and critical phenomena* (Oxford University Press, New York).

6. Plenz, D. & Thiagarajan, T. C. (2007) The organizing principles of neuronal avalanches: cell assemblies in the cortex? *Trends Neurosci* **30**, 101-110.

7. Stewart, C. V. & Plenz, D. (2006) Inverted-U profile of dopamine-NMDA-mediated spontaneous avalanche recurrence in superficial layers of rat prefrontal cortex *J. Neurosci.* **26**, 8148-8159.

8. Beggs, J. M. & Plenz, D. (2003) Neuronal avalanches in neocortical circuits *J. Neurosci.* **23**, 11167-11177.

9. Gireesh, E. D. & Plenz, D. (2008) Neuronal avalanches organize as nested theta- and beta/gamma-oscillations during development of cortical layer 2/3 *Proc. Natl. Acad. Sci. U. S. A* **105**, 7576-7581.

10. Petermann, T., Thiagarajan, T., Lebedev, M. A., Nicolelis, M. A., Chialvo, D. R. & Plenz, D. (2009) Spontaneous cortical activity in awake monkeys composed of neuronal avalanches *Proc. Natl. Acad. Sci. U. S. A* **106**, 15921-15926.

11. Chen, D.-M., Wu, S., Guo, A. & Yang, Z. R. (1995) Self-organized criticality in a cellular automaton model of pulse-coupled integrate-and-fire neurons *J. Phys. A:Math. Gen.* **28**, 5177-5182.

12. Eurich, C. W., Herrmann, J. M. & Ernst, U. A. (2002) Finite-size effects of avalanche dynamics *Phys. Rev. E. Stat. Nonlin. Soft. Matter Phys.* **66**, 066137.

13. de Arcangelis, L., Perrone-Capano, C. & Herrmann, H. J. (2006) Self-organized criticality model for brain plasticity *Phys Rev. Lett* **96**, 028107.

14. Rohrkemper, J. & Abbott, L. F. (2006) A simple growth model constructs critical avalanche networks. *Prog. Brain Res.* **165**, 13-19.

15. Pellegrini, G. L., de Arcangelis, L., Herrmann, H. J. & Perrone-Capano, C. (2007) Activity-dependent neural network model on scale-free networks *Phys. Rev. E. Stat. Nonlin. Soft. Matter Phys.* **76**, 016107.

16. Teramae, J. N. & Fukai, T. (2007) Local cortical circuit model inferred from power-law distributed neuronal avalanches *J. Comput. Neurosci.* **22**, 301-312.

17. Omori, F. (1895) On the aftershocks of earthquakes. *J. Coll. Sci. Imper. Univ. Tokyo* **7**, 111-200.

18. Utsu T., Ogata Y. & , M. R. S. (1995) The centenary of the Omori formula for a decay law of aftershock activity. *J. Phys. Earth* **43**, 1-33.

19. Shew, W. L., Yang, H., Petermann, T., Roy, R. & Plenz, D. (2009) Neuronal avalanches imply maximum dynamic range in cortical networks at criticality *J. Neurosci.* **29**, 15595-15600.

20. Stewart, C. V. & Plenz, D. (2007) Homeostasis of neuronal avalanches during postnatal cortex development in vitro *J. Neurosci. Meth.* **169**, 405-416.

21. Zapperi, S., Baekgaard, L. K. & Stanley, H. E. (1995) Self-organized branching processes: Mean-field theory for avalanches *Phys. Rev. Lett.* **75**, 4071-4074.

22. Levina, A., Hermann, J. M. & Geisel, T. (2007) Dynamical synapses causing self-organized criticality in neural networks *Nature Physics* 857-860.

23. Kinouchi, O. & Copelli, M. (2006) Optimal dynamical range of excitable networks at criticality *Nature Physics* **2**, 348-351.

24. Sornette, D. & Knopoff, L. (1997) The paradox of the expected time until the next earthquake. *Bull. Seismol. Soc. Am.* **87**, 789-798.

25. Osorio, I., Frei, M. G., Sornette, D. & Milton, J. (2009) Pharmaco-resistant seizures: self-triggering capacity, scale-free properties and predictability? *Eur. J. Neurosci.* **30**, 1554-1558.

26. Olami, Z., Feder, H. J. S. & Christensen, K. (1992) Self-organized criticality in a continuous, nonconservative cellular automaton modeling earthquakes *Phys. Rev. Lett.* **68**, 1244-1247.

27. Bak, P., Christensen, K., Danon, L. & Scanlon, T. (2002) Unified scaling law for earthquakes *Phys. Rev. Lett.* **88**, 178501.

28. Corral, A. (2003) Local distributions and rate fluctuations in a unified scaling law for earthquakes *Phys. Rev. E.* **68**, 035102.

29. Corral, A. (2004) Long-term clustering, scaling, and universality in the temporal occurrence of earthquakes *Phys. Rev. Lett.* **92**, 108501.

30. Paczuski, M., Maslov, S. & Bak, P. (1996) Avalanche dynamics in evolution, growth, and depinning models *Phys. Rev. E.* **53**, 414-443.

31. Albano, E. V. (1995) Spreading analysis and finite-size scaling study of the critical behavior of a forest fire model with immune trees *Physica A* **216**, 213-226.

32. Pajevic, S. & Plenz, D. (2008) Efficient network reconstruction from dynamical cascades identifies small-world topology from neuronal avalanches *PLoS Comp. Biol.* **5**, e1000271.

33. Kitzbichler, M. G., Smith, M. L., Christensen, S. R. & Bullmore, E. (2009) Broadband criticality of human brain network synchronization *PLoS Comput. Biol.* **5**, e1000314.

# Supplementary Information

# Scaling properties of neuronal avalanches are consistent with critical dynamics

Dietmar Plenz & Dante R. Chialvo

Here we provide analysis that further demonstrates the size invariance of neuronal avalanches and its relation to the equivalent Omori-law described in the main text. In figure S1, we show that the dynamics of events with sizes belonging to the scale-free region of the size distribution obey the same scaling law, while events with sizes at the exponential cutoff exhibit distinct, finite-size behavior. In figure S1A, the size distribution for the representative network analyzed in figure 3B of the main text is shown. Six sizes ($s$ = 32 – 317 µV) that fall inside the scale-free region are denoted with vertical lines. In addition, one size (s = 502 µV, *broken line*) which falls outside the scaling regime and within the exponential cut-off is selected. In Figure S1C, in a format similar to figure 3B of the main text, the probabilities of an avalanche preceding and following a main avalanche of given size $s$ are plotted (*color code*). As can be seen, all foreshock and aftershock functions related to trigger sizes within the scaling region over plot correctly. On the other hand, the functions differ for trigger avalanches whose sizes fall outside the scale-free regime, i.e. s = 502 µV. The more striking difference relates to the foreshock dynamics, which shows very little activity preceding the relatively infrequent, largest avalanches (*broken orange line*). This is typical of a finite-size effect, where the limited size of the electrode array does not allow the observation of avalanches much larger than 502 µV, which would be necessary in order to relate these large avalanches to their near past. Of course, the use of even larger electrode arrays will eventually reveal a similar finite-size effect for avalanches >>502 µV.

The figure S1 also demonstrates that the foreshock and aftershock function within the scaling regime are reasonably robust for various considerations of inter-avalanche times with regard to the avalanche lifetimes. In figure S1B, the lifetime distribution P($T$) for the network in $A$ is plotted demonstrating that for most avalanches the lifetime $T$ is smaller than 40 ms. The *inset* demonstrates that avalanches of similar size $s$ vary tremendously in $T$, although there is a tendency for large avalanches to have long lifetimes. To understand how the lifetimes might affect the power law in the foreshock and aftershock distributions, we analyzed 4 different conditions. Comparing foreshock distributions, we see that the distributions don't change significantly whether time is measured from the end (Fig. S1C) or start (Fig. S1D) of the foreshock to the start of the mainshock. The main difference lies in the extension of the law closer to the start of the mainshock when foreshock end times are considered. Similarly, when aftershock times are calculated from the end of the mainshock to the begin of the aftershock (Fig. 1C), the function covers times immediately following the mainshock. Measuring aftershock times from the beginning of the mainshock, as shown in Fig. S1D, demonstrates the increasing influence of the lifetime particularly for large mainshocks when estimating the aftershock law.

In Figure S2 we present the null-hypothesis for the expected relationships between main avalanches and their corresponding foreshock or aftershock probabilities as well as sizes.

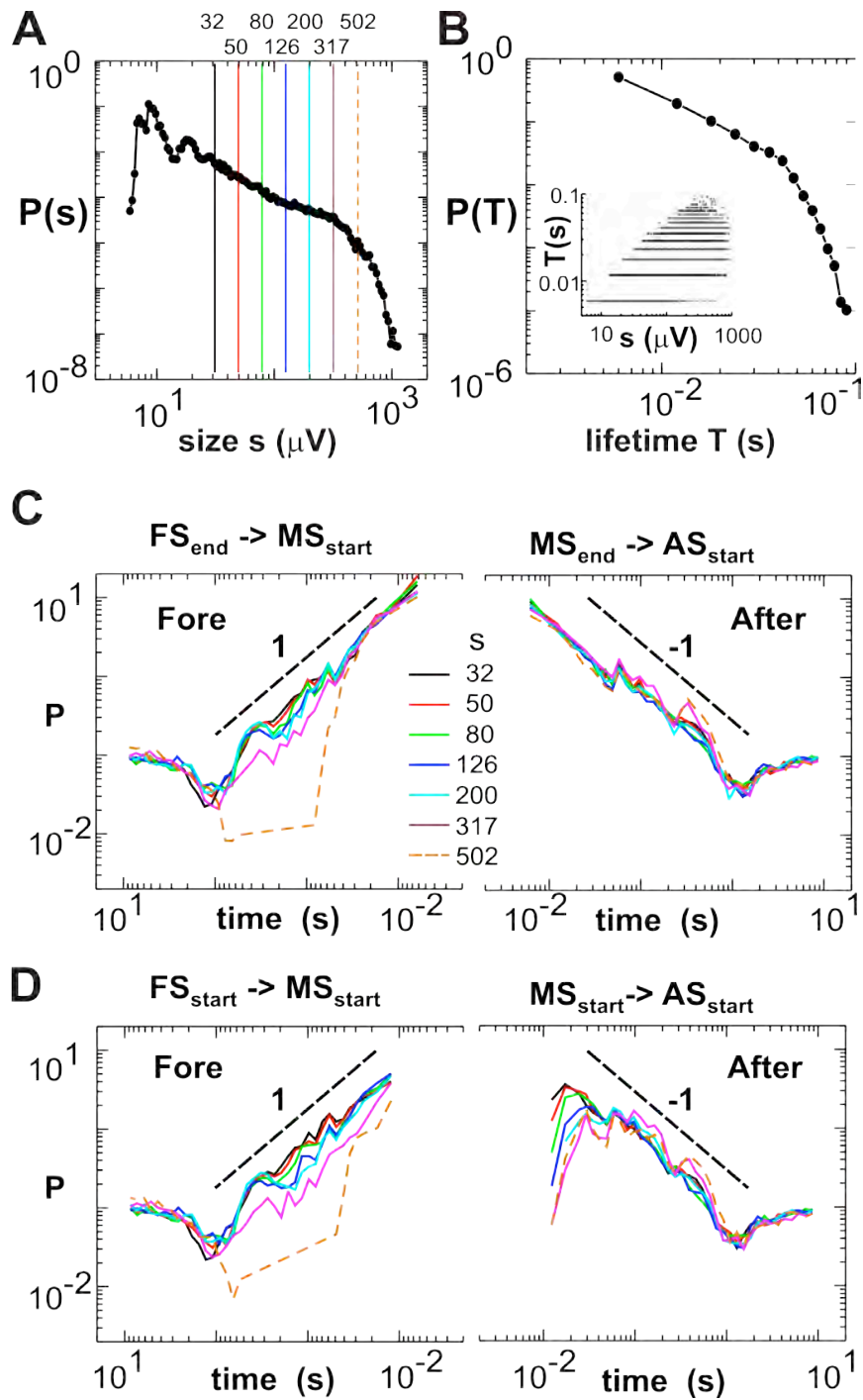

Figure S1

**Figure S1 Scale-invariance and finite-size effects in the foreshock and aftershock times of avalanches.** (*A*) Avalanche size distribution for a single representative network (cp. Fig. 3A, main text). Colored lines and numbers indicate sizes used to calculate fore and aftershock distributions in *C* and *D*. Broken line indicates size outside the scaling region. (*B*) Lifetime distribution of the network in A. Note that $T < 40$ ms for most avalanches. *Inset*: Scatterplot of size *s* and corresponding lifetime *T* for each avalanche. (*C*) Foreshock (*left*) and aftershock (*right*) distributions calculated for different avalanches of size *s*. Note that only foreshock distributions within the scaling regime ($s < 502$ µV) overlap. Main avalanches outside the regime ($s > 502$ µV) show little foreshock activity (broken line). For foreshocks, time is measured from the end of the foreshock ($FS_{end}$) to the start of the mainshock ($MS_{start}$). Conversely, for aftershocks, time is measured from the end of the mainshock (MS$_{end}$) to the start of the aftershock (AS$_{start}$). (*D*) Foreshock and aftershock distributions when time is measured from FS$_{start}$ to MS$_{start}$ and MS$_{start}$ to AS$_{start}$ respectively. Note the increasing influence of mainshock lifetime for large mainshocks on the initial part of the aftershock function.

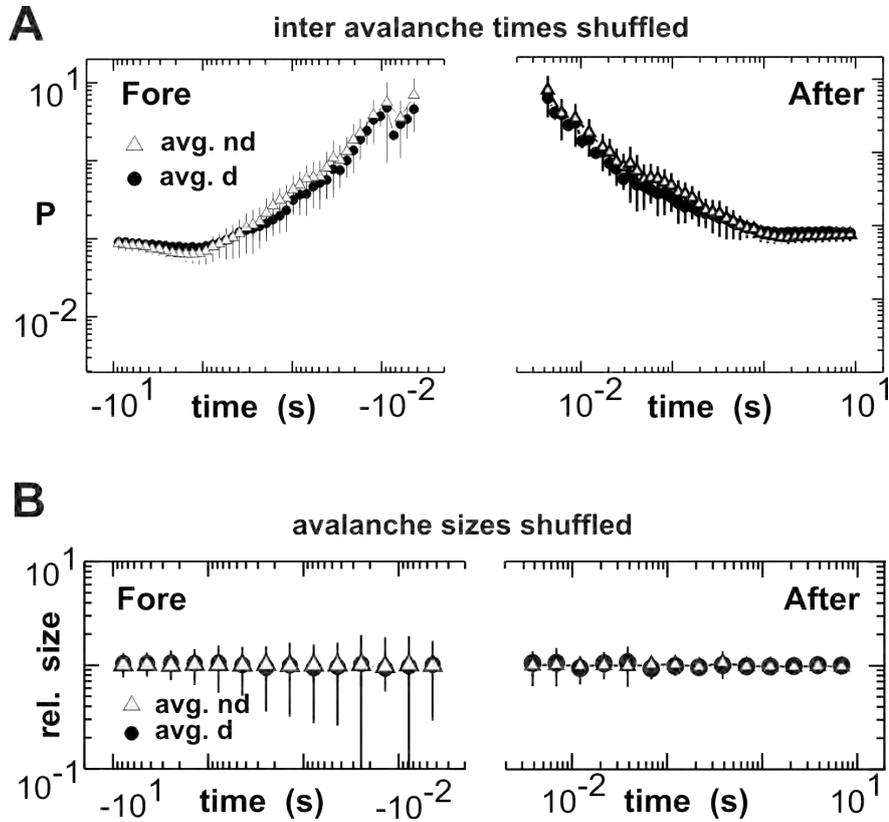

Figure S2

**Figure S2 Shuffle-controls regarding the probability of occurrence and the expected avalanche sizes before and after a main avalanche.** (*A*) Fore and after-shock probabilities of neuronal avalanches relative to the mainshock are robust to shuffling of waiting times and avalanche sizes, in both driven and non-driven networks. (*B*) The null hypothesis computed for the decay of after shock sizes by randomly shuffling avalanche sizes. This procedure removes the time dependence of after- and foreshock sizes (Average from 100 realizations at $s_0$ = 800 µV for nd and d networks).